\documentclass[a4paper]{jpconf}
\usepackage{graphicx}
\usepackage{amsmath,amssymb}
\usepackage[LGR, T1]{fontenc}
\usepackage{babel}%[polytonicgreek,english]

\usepackage{braket}
\usepackage{cite}
\usepackage{orcidlink}

\def\Ob{\mathcal{O}}

\newcommand{\I}{\mathrm{i}}
\newcommand{\D}{\mathrm{d}}
\newcommand{\eq}[1]{\begin{equation}\begin{aligned}#1\end{aligned}\end{equation}}

\def\Del#1#2{\frac{\D #1}{\D #2}}
\def\eref#1{(\ref{eq:#1})}

\def\Hc{\mathcal{H}}

\usepackage{xcolor}

\definecolor{liste}{rgb}{0,0.5,0} %dunkelgruen

\usepackage{hyperref}

\begin{document}
\title{Time and its arrow from quantum geometrodynamics?}

\author{Claus Kiefer$^{1}\orcidlink{0000-0001-6163-9519}$, Leonardo Chataignier$^{2}\orcidlink{0000-0001-6691-3695}$, and Mritunjay Tyagi$^{3}\orcidlink{0009-0002-6235-8297}$}

\address{$^1$Faculty of Mathematics and Natural Sciences, Institute for Theoretical Physics, University of Cologne, Cologne, Germany}
\address{$^2$Department of Physics and EHU Quantum Center, University of the Basque Country UPV/EHU, Barrio Sarriena s/n, 48940 Leioa, Spain}
\address{$^3$University of Groningen, University College Groningen, Hoendiepskade 23/24, 9718 BG Groningen, The Netherlands}

\ead{kiefer@thp.uni-koeln.de, leonardo.chataignier@ehu.eus,
  m.tyagi@rug.nl}

\vskip 2mm

\begin{abstract}
We discuss how quantum geometrodynamics, a conservative approach to
quantum gravity, might explain the emergence of classical spacetime
and, with it, the emergence of classical time and its arrow from the
universal quantum state. This follows from a particular but reasonable
choice of boundary condition motivated by the structure of the
Hamiltonian of the theory. This condition can also be seen as defining
a quantum version of Penrose's Weyl curvature hypothesis. We comment
on the relation of this picture to the `past hypothesis' and the
different observed arrows of time, and we consider how quantum
geometrodynamics could serve as a unifying and more fundamental
framework to explain these observations. 

\end{abstract}

\section{Introduction}
Although most fundamental laws of Nature, as we currently know them,
are time symmetric, our observations clearly distinguish a
``direction'' or ``arrow'' of time. This becomes evident from
different types of phenomena: the propagation of radiation (e.g. the
distinction between advanced and retarded potentials); the expansion
of the Universe and the formation of structure, including black holes
(both related to the gravitational field); the general tendency of
entropy to increase, as explained by thermodynamics and statistical
physics; and the apparent asymmetry and irreversibility of quantum
measurements (apparent ``collapse'' of the wave function). Could there
be a ``master arrow'' of time behind these different phenomena? In
other words, could there be a single, well-motivated reason for all
these different arrows? 

In this article, we entertain the possibility that, once the quantum
nature of the gravitational field and the fundamental ontology of
quantum mechanics are well understood, then time and its master arrow
may follow from the structure of the quantum state itself. In a
nutshell, this is due to the fact that, as we currently know of no
bounds to the quantum superposition principle, and given the ubiquity
of entanglement, it is possible that the whole Universe is described
by a single quantum state as a closed quantum system. Furthermore, as
the prevailing interaction at large scales is gravitation, the
description of the Universe as a single quantum system ought to
include a quantum description of gravity, even if we aim to describe
only a reduced set of large-scale degrees of freedom instead of
literally everything. Finally, due to spacetime diffeomorphism
invariance, the universal quantum state must be independent of any
time coordinate, so any physical notion of time and its arrow must be
`internal' (or `intrinsic'); that is, it must emerge from the quantum
degrees of freedom themselves. 

In this way, we are led to the prospect that the Universe could be
fundamentally described by a quantum state, from which dynamics and
history emerge. The general outline of the ideas reported here have
been previously discussed in
\cite{KieferDice,ZehBook,KieferArrows,ZehArrows,Conradi,KieferWeyl}. Although there are other
formalisms (see, for instance, the different contributions to the book
in \cite{KieferArrows,ZehArrows}), it is important to emphasize how a
conservative approach to the canonical quantization of the
gravitational field, the theory of quantum geometrodynamics, which
only assumes general relativity together with the usual tenets of
quantum theory and nothing more, can consistently point us toward an
explanation of the emergence of classical time and its arrow. 

This contribution to the Symposium proceedings consists of six sections. We
first review the various arrows of time and the idea of the {\em past
  hypothesis}.\footnote{A general reference is the textbook
  \cite{ZehBook}, to which we refer for more details and references to
  original literature.} We then present a brief section where we focus
on the situation in quantum theory. Section~4 and~5 constitute the
main part of our contribution. In section~4 we present the main ideas
on how the arrows of time can emerge from the fundamental framework of
quantum geometrodynamics, and section~5 contains the outline of an example for which technical details and further developments are relegated to a separate publication. We conclude in section~6.

\section{Arrows of time and the past hypothesis}
We observe many phenomena that frequently occur in a certain order but
infrequently in the opposite order. This discrepancy defines an arrow
of time as Eddington explained \cite{EddingtonArrow} and it brings
forth an element of irreversibility into our description of
Nature. Historically, as most of the physical laws that we have
understood are time symmetric,\footnote{There are, of course,
  phenomena associated with the weak interaction that are not time
  symmetric. Nonetheless, they are still CPT invariant, which can be
  interpreted as a slight generalization of $T$-invariance,
  drastically different from the irreversibility discussed here.}
the origin of irreversibility has been a vexing issue (Loschmidt's
{\em Umkehreinwand} and Zermelo's {\em
  Wiederkehreinwand}; see e.g. \cite{ZehBook,WuZermelo}).

One of the most prevalent arrows of time is the thermodynamical arrow, which is responsible for the time directionality of many events we observe. The second law of thermodynamics establishes the general tendency of entropy to increase:
\eq{
\frac{\D S}{\D t}= \underbrace{\left(\frac{\D S}{\D t}\right)_{\rm ext}}_
 {\D S_{\rm ext}=\delta Q/T} +\underbrace{\left(\frac{\D S}{\D t}
    \right)_{\rm int}}_{\geq 0} \ ,
}
and this growth of entropy for an isolated system defines the thermodynamical arrow of time. Its relevance notwithstanding, the question of its fundamental physical origin remains.

While it is clear that the direction of time is related to the second
law of thermodynamics, it is not clear whether this is the master
arrow (if it exists). Over the years, the precise origin of time's
directionality has been a topic of continued debate. For example,
Einstein and Ritz represented different opinions regarding the matter
in 1909. In a common paper, they state
\begin{quote}
    `` \ldots Ritz considers the restriction to the form of the retarded
potentials as one of the roots of the Second Law, while Einstein
believes that the irreversibility is exclusively based on reasons of
probability.''\footnote{\scriptsize German original: ``Ritz betrachtet
  die Einschr\"ankung auf die Form der retardierten Potentiale als
  eine der Wurzeln des Zweiten Hauptsatzes, w\"ahrend Einstein glaubt,
  da\ss\ die Nichtumkehrbarkeit ausschlie\ss lich auf
  Wahrscheinlichkeitsgr\"unden beruhe.''}
\end{quote}
Indeed, in electrodynamics, it is clear that fields interacting with
local sources are usually described by 
retarded solutions. Since a general solution of the wave equation can
be decomposed as \cite{ZehBook}
\eq{
\mbox{four-potential} = A^{\mu}&= \mbox{source term plus boundary term}\\
       &= A^{\mu}_{\rm retarded}+ A^{\mu}_{\rm in}\\
       &= A^{\mu}_{\rm advanced}+ A^{\mu}_{\rm out} \ ,
}
the restriction to retarded potentials corresponds to the choice of
the `Sommerfeld
condition' $A^{\mu}_{\rm in}\approx 0$. Why should such a
condition hold? 

While Ritz essentially expressed the opinion that the electrodynamic
(or radiative) arrow could be a candidate for a master arrow of time
(which one could perhaps see as a fundamental fact of Nature),
Einstein's view is the one favoured today, as it alluded to a
statistical explanation of the second law of thermodynamics as the
more fundamental concept, with which other arrows, including the
electrodynamic one, could conceivably be explained. In fact, the
electrodynamic arrow can be understood as a consequence of the second
law and the thermodynamic arrow by taking into account thermodynamic
properties of absorbers. 

The statistical foundations of thermodynamics, which were most notably
developed by Boltzmann and Gibbs, attribute the non-decreasing
character of the entropy of an isolated system to what Einstein calls
``reasons of probability.'' Low-entropy states are assumed to be less
probable than high-entropy states, and thus one expects that states with
lower entropy will evolve into states with higher entropy. If this
reasoning is extended to the whole Universe, then the observed growth
of entropy indicates that entropy was low in the early Universe. This
would thus be an initial or boundary condition for the evolution of
the Universe. According to Boltzmann (1898): 
\begin{quote}
    That in Nature the transition from a probable to an improbable
    state does not happen equally often as the opposite transition,
    should be sufficiently explained by the assumption of a very
    improbable initial state of the whole Universe surrounding us
    (\ldots), 
\end{quote}
or Arthur Eddington \cite{Eddington1931}:
\begin{quote}
Accordingly, we 
sweep anti-chance [low-entropy state generated by an extremely
improbable fluctuation] out of the laws of physics---out of the differential
equations. Naturally, therefore, it reappears in the boundary
conditions (\ldots).
\end{quote}
This boundary condition is often referred to as the ``past
hypothesis'' (see \cite{AlbertPastHyp,PastHypBook} and also the
discussion in \cite{Earman}), which would explain the arrows of time
based on a fundamental assumption about the early
Universe.\footnote{Note that the meaning of `past' is itself defined
  by this hypothesis.} But is
that really the case? 

Similarly to the situation in eletrodynamics, arrows of time arise
also in gravitation and cosmology. Indeed, the formation of structure
in the Universe (such as stars, galaxies, and clusters), which follows
from gravitational instabilities, and the expansion of the Universe
can be thought of as defining gravitational and cosmological arrows of
time. Can they be connected to the second law of thermodynamics? If
so, one must have a notion of entropy of the gravitational degrees of
freedom. As gravitational systems have a negative heat capacity,
(in)homogeneous states are to have (high) low gravitational entropy,
the opposite behaviour with respect to non-gravitational
systems.

As gravitational radiation and tidal forces can be captured by the
Weyl curvature tensor, it is reasonable to expect that the entropy of
gravitational fields may be related to Weyl curvature. Roger Penrose
proposed the `Weyl curvature hypothesis' \cite{PenroseWCH} in order to
suggest a possible explanation for the homogeneity and isotropy of the
observed Universe at large scales, as well as for the growth of
entropy in the Universe. This hypothesis states that the Weyl
curvature tensor must vanish (or at least be non-divergent) at past
singularities approached from the future. With this, gravitational
waves must be retarded, in a direct analogy to the Sommerfeld
condition for electromagnetic waves. In contrast to the electrodynamic
case, it is difficult to relate the gravitational and cosmological
arrows to the thermodynamic one due to the weakness of gravitational
waves.

A vanishing Weyl curvature tensor in the early Universe is thus
expected to be related to the Universe's initial low entropy. As the
Universe evolved, the growth of entropy, the arrow of time, and the
formation of structure would have followed from the increasing effect
of Weyl curvature on the dynamics. Although it remains to be seen
whether the Weyl curvature hypothesis is correct, the effort to take
into account the entropy of gravitational fields is a step in the
direction of unifying the arrows of time under a common entropic
explanation based on the past hypothesis. 

It is also worthwhile to mention that the Bekenstein--Hawking entropy
of black holes, 
\eq{
S_{\rm BH}=k_{\rm B}\frac{Ac^3}{4\hbar G}\stackrel{\rm Schwarzschild}
{\approx} 1.07\times 10^{77}k_{\rm B}
\left(\frac{M}{M_{\odot}}\right)^2 \ ,
}
indicates the importance of entropy in gravitation, especially at its
interface with quantum field theory, and it begs a microscopic
explanation (as several approaches to quantum gravity aim to yield
\cite{KieferBook}). With this analysis, one can also estimate the
``probability of our Universe'' \cite{KieferArrows,PenroseProb}, by
considering that the maximal entropy of the observed Universe (more
precisely: its region within the particle horizon) is
obtained if its mass is concentrated in a single black hole. The
fantastical result\footnote{In this estimate, we also include the
  Gibbons--Hawking entropy \cite{GHEnt} associated with $\Lambda$
  \cite{KieferArrows}.} \cite{KieferArrows,PenroseProb} 
\eq{
\frac{\exp\left(\frac{S}{k_{\rm B}}\right)}{\exp\left(\frac{S_{\rm
      max}}{k_{\rm B}}\right)}\sim \frac{\exp\left(3.1\times10^{104}\right)}
 {\exp\left(2.9\times 10^{122}\right)}
\approx \exp\left(-2.9\times 10^{122}\right) \label{universe-entropy}
}
indicates how special the initial state must be. In general, this
might raise issues of fine tuning, although we will see how toy models
of quantum gravity could address the problem via a reasonable choice
of boundary condition, which implements a version of the past
hypothesis.

The number \eqref{universe-entropy} is an unimaginably tiny number. In
fact, it is much smaller than the probability of the whole observable
Universe including all states of observers emerging in a gigantic
fluctuation in the spirit of Boltzmann. This number would correspond
to the total entropy of the Universe and thus be of the order
$10^{-104}$ -- still tiny but 18 orders of magnitude bigger than
$10^{-122}$. This discrepancy casts strong doubts on the applicability
of the anthropic principle, as the smallness of
\eqref{universe-entropy} calls for a physical explanation.

\section{Irreversibility and entropy in quantum theory}
Even though the Schr\"odinger equation is invariant under time
reversal, the measurement process distinguishes a direction of time
via the update (or ``collapse'') of the wave function, which might be
ontic (e.g., a dynamical collapse) or epistemic. In
particular, we can take the Everettian view that the quantum state and
the Schr\"odinger equation are the ontological elements of the theory,
from which all else is to be derived. In this view, the measurement
process entails the branching of Everett paths (the ``many
worlds''). This is certainly economical from the point of view of
formalism, although some (like John Bell) may consider the worlds an ontological
extravagance. As we wish to make no further modifications or additions to the
formalism here, we adopt the Everett point of view; it is certainly
the point of view implicitly assumed in many discussions of quantum
cosmology. 

A central concept related to the process of measurement and the
branching of the quantum state is decoherence, which follows from the
increasing entanglement that results from interactions with the
environment, and it leads to the irreversible emergence of classical
properties \cite{DecoBook}. Can this quantum irreversibility and the
corresponding arrow of time be related to a definition of entropy? 

The increasing entanglement can be quantified by the definition of
entanglement entropy, for example via the von Neumann or linear
entropies, respectively given by 
\eq{
S_{\rm vN}&=-k_{\rm B}{\rm tr}\left(\rho\ln\rho\right)\\
S_{\rm lin} &= k_{\rm B}{\rm tr}\left(\rho-\rho^2\right) \ ,
}
where $k_{\rm B}$ is Boltzmann's constant (needed to express the
entropy in thermodynamic units) and $\rho$ is the density matrix
representation of the quantum state. This density matrix is reduced in
the sense that the environment degrees of freedom are to be traced
out, which corresponds to a type of coarse graining, and it leads to a
mixed state if the system is entangled with the environment. In
general, coarse graining entails the transformation of relevant
information (e.g. about macroscopic differences) into irrelevant
information (e.g. about microscopic differences without discernible
macroscopic consequences),\footnote{The classic example is Gibbs's ink
  drop analogy.} and this may lead to an increase in
entropy (from zero for a pure state to nonzero for a mixed state).  

Although the precise relation between entanglement entropy and
thermodynamical entropy must be spelled out, it is conceivable that,
if quantum theory is fundamental and describes the whole Universe,
entanglement entropy may play a key role in establishing and
explaining a master arrow of time, from which the other arrows (e.g.
thermodynamical, gravitational, electrodynamic, cosmological) may
emerge. Motivated by this open question, we are led to consider the
question of quantum gravity: if the whole Universe is a single quantum
system that includes a quantum description of gravitation, what is the
role of entanglement entropy and the past hypothesis? Is it possible
that quantum gravity accommodates (or ideally explains) the past
hypothesis and the observed arrows of time? 

In what follows, we would like to emphasize that a conservative
approach to quantum gravity, which only assumes the usual tenets of
general relativity and (Everettian) quantum theory without further
speculative elements, may point towards an explanation of the
emergence of classical time and its master arrow from a fundamental
theory in which the quantum state and the (time-independent)
Schr\"odinger equation are the basic ontological elements. This
envisioned explanation is of course tentative because the theory of
canonical quantum gravity is not yet fully developed and, moreover,
the connection between entanglement entropy and its other counterparts
must still be properly understood, as mentioned above. 

\section{Quantum gravity, the timeless past hypothesis, and quantum
  Weyl curvature} 
In canonical quantum gravity, which was pioneered by Rosenfeld,
Bergmann, and Dirac, and developed by Wheeler and DeWitt and others
\cite{KieferBook}, 
the gravitational field is quantized according to the usual rules of
canonical quantization, and the result is a (functional) Schr\"odinger
equation for gravity and matter fields. Due to spacetime
diffeomorphism invariance, the quantum states must be invariant under
time translations, and this implies that the fundamental Schr\"odinger
equation must be time independent (or timeless): 
\eq{\label{eq:WDW0}
0 = \I\hbar\Del{}{\tau}\ket{\Psi} = \hat{H}\ket{\Psi} \ .
}
This equation is better known as the Wheeler--DeWitt
equation. The timelessness of this constraint on the
physical states simply signals that no external or absolute time has
physical meaning in the theory, and thus a physical notion of
evolution, if any, must be internal, that is, derived from the
physical degrees of freedom themselves. The intuitive explanation for
this timelessness is that there is no time ``outside'' spacetime in
general relativity, and that the classical spacetime is absent in the
quantum theory in the same way as a classical trajectory is absent in
quantum mechanics; see, for example, \cite{DeWitt67,KieferBook,CKPP} for more details.  

The quantum state $\ket{\Psi}$ leads to the wave function
$\Psi(g_{\mu\nu},\phi)$, which in general depends on the spacetime
metric $g_{\mu\nu}$ and on matter fields $\phi$. If we perform a $3+1$
decomposition of these variables (which corresponds at the classical
level to a foliation of
the four-dimensional spacetime into spacelike leaves $\Sigma$), then
the requirement of diffeomorphism invariance is 
equivalent to the validity of the local quantum constraints 
\eq{
\hat{\Hc}_{\perp}\Psi = 0 \ , \ \hat{\Hc}_{i}\Psi = 0 \ , 
}
where $\hat{\Hc}_{i}$ generates spatial diffeomorphisms,
$\hat{\Hc}_{\perp}$ is responsible for time reparametrizations, and
the Hamiltonian operator is $\hat{H} = \int_{\Sigma}\D^3 x
N^{\mu}\hat{\Hc}_{\mu}$ with
$
\mu=(\perp,i=1,2,3)$, and $N^{\mu} = (N^{\perp}\neq0,N^i)$ being arbitrary
functions. After imposing all constraints, it is seen that the wave
function only depends on the {\em spatial} metric, whose components are
here called $h_{ab}$.

If $\Psi(h_{ab},\phi)$ only depends on combinations of
the fields that are invariant under spatial diffeomorphisms, then the
constraint $\hat{\Hc}_{i}\Psi = 0$ is satisfied. The remaining
Hamiltonian constraint reads (for
simplicity, the matter content is that of a minimally coupled
scalar field) \cite{KieferBook,DeWitt67}:
\eq{\label{eq:WDW}
0&=\sqrt{h}\hat{\Hc}_{\perp}\Psi=\left[-\frac{16\pi G\hbar^2}{c^2}\left(h_{ac}h_{bd}-\frac12h_{ab}h_{cd}\right)\frac{\delta^2}{\delta h_{ab}\delta h_{cd}}-\frac{\hbar^2}{2}\frac{\delta^2}{\delta\phi^2}+V\right]\Psi \ , \\
V &= -\frac{c^4}{16\pi
  G}h(R-2\Lambda)+hh^{ab}\phi_{,a}\phi_{,b}+h\mathcal{V}(\phi) \ , 
}
where $G$ is Newton's constant, $h$
is the determinant of the spatial metric, $R$ is the Ricci scalar on
the spatial leaves, 
$\Lambda$ is the cosmological constant, and $\mathcal{V}(\phi)$ is an
arbitrary potential for the scalar field. We have ignored factor
ordering ambiguities in Eq. \eref{WDW}. 

Notice how the potential term $V$ in Eq. \eref{WDW} vanishes in the
limit $h = \det h_{ab}\to0$, thus yielding an approximate separable
equation for gravity and matter, which admits separable solutions [see
Eq. \eref{SBC} below]. This signals a fundamental asymmetry of the
Hamiltonian constraint with respect to the ``local volume'' factor
$\sqrt{h}$.\footnote{More precisely, the potential term $V$ in Eq.~\eref{WDW} generally changes under the substitution $h\to-h$.} The Schr\"odinger equation of canonical quantum gravity
thus distinguishes a direction in configuration space,\footnote{The local volume factor is also distinguished by the signature of the DeWitt
metric on the gravitational configuration space, also known as ``superspace''.} and we will see how this
is related to a distinguished direction of classical time. 

Although the definition of the Hamiltonian $\hat{H}$ in the full
theory requires regularization, and the solutions to Eqs. \eref{WDW0}
and \eref{WDW} are not readily available, one can already begin to
understand the phenomenology of the Wheeler--DeWitt equation in simple
toy models which do not aim at describing all possible degrees of
freedom, but rather a subset of variables that are relevant to the
description of, for instance, degrees of freedom relevant to
cosmology. We are thus led to the topic of quantum cosmology, where
the quantum state $\ket{\Psi}$ leads to the wave function of the
universe,\footnote{A universe (with lower-case u) is a toy model of the
  real Universe.} which is in line with the notion that quantum
theory is universal.\footnote{Possible observational tests of quantum
  cosmology are discussed, for example, in \cite{CKM}; see also references therein.}

The Hamiltonian constraint in simple
toy models of quantum cosmology (assuming a homogeneous and isotropic
universe with small perturbations) can be shown to be \cite{Conradi,KieferDice,ZehBook,KieferArrows,ZehArrows,KieferWeyl}: 
\eq{\label{eq:structure-wdw}
\hat{H}\Psi=\left[\frac{\partial^2}{\partial\alpha^2}
+\sum_i\left(-\frac{\partial^2}{\partial x_i^2}+
\underbrace{V_i(\alpha,x_i)}_{\to 0 \ 
\mbox{for}\ \alpha\to -\infty}\right)\right]\Psi=0 \ ,
}
where $\alpha=\log(a/a_0)$ is the logarithm of the scale factor of the
universe ($a_0$ is a reference scale factor) and $x_i$ are other
degrees of freedom. Also here, the 
potential terms vanish as $a\to0$ or $\alpha\to-\infty$ (the analogous
condition of the general case $h\to0$). In this way, we can consider
the boundary condition
\cite{Conradi,KieferDice,ZehBook,KieferArrows,ZehArrows,KieferWeyl} 
\eq{\label{eq:SBC}
\Psi\stackrel{\alpha\to
  -\infty}{\longrightarrow}\psi_0(\alpha)\prod_i\psi_i(x_i) \ .
}
As the region of configuration space in which $a\to0$ corresponds, in
particular, to the early universe, such a symmetric boundary condition
corresponds to the imposition of a completely unentangled state ``in
the beginning.''\footnote{Conradi and Zeh compare such a symmetric
  ``initial'' condition with the {\em apeiron}
  (\foreignlanguage{polytonicgreek}{ἄπειρον}) discussed by 
  Anaximander of Milet in about 550~BC \cite{Conradi}.} This leads to
a vanishing ``initial'' entanglement entropy. If such an entropy could
also be connected to other definitions (in particular to
thermodynamical entropy \cite{Peres}), then this condition would correspond to a
timeless version of the past hypothesis, as there is no external,
absolute nor preferred time in this context. It is interesting that,
while this boundary condition is still an assumption, it is a
reasonable choice allowed by the very structure of the Hamiltonian,
which lends some credibility and motivation to this version of the
past hypothesis.  

For $a>0$, the nonvanishing potential will generally lead to an
entangled state, as it will no longer be possible to separate matter
and gravitational variables. Thus, in regions of configuration space
that correspond to the ``late-time'' universe, the entanglement
entropy will no longer be zero, as it increases with increasing scale
factor. It is worthwhile to emphasize that this can be seen as a
timeless (static) situation, and the variation of entanglement entropy
with $a$ concerns variations in the form of the wave function across
configuration space. Classical time itself, and our usual classical
notions of causality and dynamics, can be defined in the regions of
configuration space where decoherence ensures that a classical
spacetime background emerges
\cite{ZehTime,KieferContinuousMeasurement,KieferBook}.\footnote{We are
  tacitly assuming a closed universe. In the presence of an asymptotic
  structure (e.g. asymptotic flatness), a notion of time is available
  at the boundary and may persist even in the quantum theory, while
  the constraints (in particular, the Wheeler--DeWitt equation) are
  valid in the bulk, where the emergence of a classical background
  through decoherence can be studied.} Here, the system includes the
background degrees of freedom, and possibly some of the Weyl scalars
that describe scalar and tensor modes, whereas the environment is, for example,
composed of small density fluctuations and weak gravitational
waves. In contrast to the usual decoherence process in quantum 
mechanics, which unfolds relative to an external time, such a
quantum-cosmological decoherence responsible for the emergence of classical
time unfolds relative to the scale factor itself, and thus it also
concerns variations across configuration space. With this
understanding, it could be that canonical quantum gravity motivates
the ``initial'' low entropy state and the subsequent master arrow of
time. In this way, the asymmetry of the Wheeler--DeWitt equation and
decoherence (caused by increasing entanglement
\cite{ZehTime,KieferContinuousMeasurement,DecoBook}) would be the most
fundamental sources of irreversibility behind the arrows of
(classical) time. 

What else can we say about the asymptotic state given in
Eq. \eref{SBC}? Can we fix the form of each factor? As the symmetric
boundary condition is a choice, albeit well motivated, we can only fix
the form of the factors by a further hypothesis. In analogy to
Penrose's classical Weyl curvature hypothesis, we can introduce the following
quantum version of this hypothesis
\cite{KieferWeyl}: 
  \begin{quote}
{\em The quantum states for scalar variables that are related to scalar and
  tensor modes in cosmology assume the form of adiabatic vacuum states
  in a (quasi-)de~Sitter space, as the region of small scale factors
  is approached from directions of large scale factors.} 
\end{quote}
Instead of the classical requirement that the Weyl curvature
should vanish or that it should be at least non-divergent in the early
universe, one requires that (Weyl) scalars be in their adiabatic vacuum
states in an inflationary quantum universe, where early times are
defined by the region of configuration space with small scale
factors. This has a direct consequence to phenomenology, as
gravitational waves can be described by certain Weyl scalars
constructed from the Weyl curvature tensor \cite{NewmanPenrose}, such
as $\Psi_4:= -\frac{1}{8c^2}\left(\ddot{h}_+-{\rm
    i}\ddot{h}_{\times}\right)$,\footnote{The variable $\Psi_4$ is not to be confused with the wave function $\Psi$ that satisfies Eqs.~\eref{structure-wdw} and \eref{SBC}.} where $h_+$ and $h_{\times}$ indicate
the two polarizations of weak gravitational waves (related to the
tensor modes). The Newman--Penrose variable $\Psi_4$ corresponds to
the helicity state $s=-2$, whereas its complex conjugate describes
$s=+2$. Furthermore, the Mukhanov--Sasaki variable $v_{\vec{k}}$ combines the scalar perturbations of the metric and the
inflaton scalar field \cite{MS}, thus defining invariant scalar modes.

This quantum hypothesis then demands that the $x_i$ variables in
Eq. \eref{SBC}, which in this case correspond to the scalar and tensor
modes (or to the corresponding Weyl scalars), all now generically denoted by $v_{\vec{k}}$, be in the state 
\eq{
\psi_{\vec{k}}(v_{\vec{k}}) =
\mathcal{N}_{\vec{k}}\exp\left(-\frac{1}{2}\,
  \Omega_{\vec{k}}^{(0)}v_{\vec{k}}^2\right)\ , 
}
with $\Omega_{\vec{k}}^{(0)}=k$. This ``initial'' state at small scale
factors will evolve into a two-mode squeezed state, from which the
primordial power spectra for density perturbations and gravitational
waves can be deduced. Finally, it is important to mention that, once a
classical spacetime background has emerged at the onset of inflation
\cite{BKK}, the primordial fluctuations also go through the
quantum-to-classical transition with a more standard process of
decoherence relative to classical time, and they eventually become
classical stochastic variables that can be used to describe the seeds
for structure formation \cite{CKDP}.
  
Both the symmetric boundary condition (timeless past hypothesis) in
Eq. \eref{SBC} and the quantum Weyl curvature hypothesis are evidently
theoretical assumptions, which nevertheless appear to be well
motivated and to accommodate well our observations (power spectra in
cosmology and the arrow of time) \cite{CKM}. In order for this formalism to
become more robust, one would need not only to achieve a
regularization of the quantum Hamiltonian constraint in a full theory
of quantum gravity, but also to better understand the connection
between entanglement entropy and thermodynamical entropy in this
context (this includes the precise connection to all observed arrows
of time),
perhaps following Chap.~9 of \cite{Peres}. This
task is yet to be completed. 

\section{Outline of a toy model and associated challenges} 
Let us outline the calculation of the entanglement entropy from the Wheeler--DeWitt equation in a simplified toy model. Further details, discussions and developments will be the subject of a forthcoming
article \cite{CT}. The toy model consists of two scalar fields of the same mass $m$, where one of the fields $\phi_1$ acts as the ``system'' and the other field $\phi_2$ acts as the ``environment,'' with the constraint equation
\begin{equation}\label{WDW-toy-0}
    \left\{\frac{1}{m_{\rm P}^2}\frac{\partial^2}{\partial
        \alpha^2}-\frac{\partial^2}{\partial
        \phi_1^2}-\frac{\partial^2}{\partial
        \phi_2^2}+a_0^6\e^{6\alpha}\left[m^2(\phi_1^2+\phi_2^2)+g\phi_1\phi_2+m_{\rm
          P}^2\frac{\Lambda}{3}\right]\right\}\!\Psi(\alpha,\phi_1,\phi_2)\!=0 \,,
\end{equation}
and $0<g\ll m^2$. The parameters have mass dimensions $[\phi_{1,2}] =
[m] = [m_{\rm P}] = [|\Lambda|^{\frac12}] = [g^{\frac12}] = [1/a_0] =
1$ and $[\alpha] = 0$.\footnote{The inverse mass dimension of $a_0$ is
  the result of absorbing a length scale (with $c = \hbar = 1$) that
  comes from the integration of the spatial volume in this homogeneous
  model.} As it is more convenient to work with dimensionless
quantities, we can rescale a quantity $f\to m_{\rm P}^n f$ if $[f] =
n$. Furthermore, we are interested in the early Universe, in which $a
= a_0 \e^{\alpha}\to 0$. For convenience, we define our symmetric
boundary condition at $\alpha = \alpha_0$, with $\alpha_0<0,
|\alpha_0|\gg1$ (the limit $a\to0$ is thus obtained with
$\alpha_0\to-\infty$). It is then convenient to define the variable $s
= \alpha-\alpha_0$, so that the boundary condition is imposed at $s =
0$. Defining $0<\lambda = a_0^6\e^{6\alpha_0}\ll1$, the Wheeler--DeWitt equation
becomes 
\begin{equation}\label{WDW-toy-2}
    \left\{\frac{\partial^2}{\partial
        s^2}-\frac{\partial^2}{\partial
        \phi_1^2}-\frac{\partial^2}{\partial
        \phi_2^2}+\lambda\e^{6s}\left[m^2(\phi_1^2+\phi_2^2)+g\phi_1\phi_2+\frac{\Lambda}{3}\right]\right\}\!\Psi\!=0 \,,
\end{equation}
which is now written in terms of the dimensionless quantities. For vanishingly small scale factors ($\lambda = 0$),  we can neglect the potential term in Eq.~\eqref{WDW-toy-2} as mentioned before. In this case, we search for a separable solution (at $s = 0$, corresponding to our boundary condition). As the size of the universe increases (with $1\gg\lambda>0$ fixed and with increasing $s$), the potential term becomes relevant. In order to solve the constraint analytically, we consider a perturbative expansion in $\lambda$ in the form
\eq{
\Psi = \e^{\sum_{n = 0}^{\infty}(\lambda\e^{6s})^nS_n(s,\phi_1,\phi_2)} \equiv \Psi_0(1+\lambda\e^{6s}\Psi_1)+\Ob(\lambda^2) \ .
}
It is advantageous to expand only the exponent (instead of the wave function) in powers of $\lambda$. Keeping only the lowest orders, we would obtain $\Psi = \e^{S_0+\lambda\e^{6s} S_1}\e^R$, where $R$ denotes the neglected remaining corrections to the phase and amplitude. Due to $\lambda\ll1$, we can assume that, at field values that are not too large, the real and imaginary parts of the neglected corrections $R$ are much smaller than the contributions of $S_0$ and $S_1$. However, for large field values, the contributions from $R$ could conceivably be larger than those of the lowest orders, invalidating perturbation theory.

It is straightforward to verify that a possible solution is $S_0 = \I(k_1\phi_1+k_2\phi_2\pm\sqrt{k_1^2+k_2^2}s)$ (plane wave) and $S_1 = A+B\phi_1+C\phi_2-D(\phi_1^2+\phi_2^2)/2+E\phi_1\phi_2$, where the coefficients are complex functions of $k_{1,2}$ and the coupling constants. Then, one can either consider a superposition of such solutions for different values of $k_{1,2}$ (e.g., a narrow Gaussian centered at $\bar{k}_{1,2}$, which leads to a separable solution at $s = 0$ and $\lambda = 0$) or one can simply fix the values of $k_{1,2}$. For example, since the constraint is invariant under the exchange of $\phi_1$ and $\phi_2$, it is convenient to fix $k_1 = k_2$ to obtain a Gaussian solution that is symmetric under this exchange (this solution is also separable at $\lambda = 0$). If this solution is to be normalizable (even non-perturbatively in $\lambda$), we assume that $|\Psi|^2$ vanishes as $|\phi_{1,2}|\to\infty$. This then entails that the remainder term $e^R$ cannot diverge or grow faster than the Gaussian decreases as $|\phi_{1,2}|\to\infty$, and thus the higher-order corrections can be neglected also for very large values of $|\phi_{1,2}|$ in this case, and we can simply approximate the state as the Gaussian $\Psi = \e^{S_0+\lambda\e^{6s} S_1}$. This solution could then be normalized upon integration over $\phi_{1,2}$ (for fixed $s$), which would lead to a linear entropy of the form:
\begin{equation}
    S_\text{lin} \propto g\lambda\e^{6s}+\ldots
\end{equation}
with $\phi_2$ taken as the environment field, and the entropy would be evaluated at the lowest non-trivial order of the coupling constants and $\lambda$, with a positive proportionality constant. This has the desired behavior: the entropy vanishes as $\lambda\to0$ (low ``initial'' entropy in the ``early Universe'') and it increases with respect to $s$ (or with respect to the scale factor, thereby increasing with the expansion of the universe).

However, this result is necessarily tentative. First, it makes use of
a perturbative expansion, the validity of which needs to be thoroughly
verified. Second, one must ascertain how sensitive the result is to
the choice of state $\Psi$. Third, the entropy is obtained by manually
normalizing the state, without considering in detail the definition of
the inner product. As other works show \cite{CKM}, the
positive-definite inner product may acquire a non-trivial form even in
the matter sector when we use techniques from canonical gauge
theory. Furthermore, the precise form of the inner product has
implications for decoherence, which is another important ingredient to
the emergence of a classical background with a notion of time such as
$s$. Thus, this is only a first, but promising attempt at a concrete study of the growth of entropy in the geometrodynamical account of quantum cosmology.

\section{Conclusions and outlook}
To perform rigorous calculations for more realistic models is a challenge for future works, which might also require the use of sophisticated numerical methods. We believe, however, that the procedure outlined above could point towards a physical explanation of the growth of entropy from the conservative approach of quantum geometrodynamics, a timeless theory that only assumes general relativity and the principles of quantum theory. This is in line with recent work \cite{KC} that discusses the ``entanglement past hypothesis.'' As we have seen, this hypothesis can be accommodated by a reasonable choice of boundary condition related to the structure of the Hamiltonian in general relativity. An open issue is, of course, whether the specification of this boundary condition as expressed by the quantum Weyl tensor hypothesis can be derived from an underlying principle instead of being freely chosen. These and other questions motivate further research in this topic.

\vspace{-0.25cm}
\ack{C.K. thanks the organizers of the {\em Symmetries in
  Science Symposium~2023} for giving him the opportunity to present a talk
  on this topic. The work of L.C. is supported by the Basque Government Grant
  \mbox{IT1628-22}, and by the Grant PID2021-123226NB-I00 (funded by
  MCIN/AEI/10.13039/501100011033 and by ``ERDF A way of making
  Europe''). It is also partly funded by the IKUR 2030 Strategy of the
  Basque Government.}

\end{document}